Dear dr. Dremin,

   please find included below the latex file of my contribution.
I'm sorry for the delay. 

                                Sincerely yours,

                                 Roberto Turolla

\documentstyle[sprocl]{article}

\def\Journal#1#2#3#4{{#1} {\bf #2}, #3 (#4)}

\def\be{\begin{equation}}
\def\ee{\end{equation}}
\def\bea{\begin{eqnarray}}
\def\eea{\end{eqnarray}}
\def\mincir{\raise -2.truept\hbox{\rlap{\hbox{$\sim$}}\raise5.truept
\hbox{$<$}\ }}
\def\magcir{\raise -4.truept\hbox{\rlap{\hbox{$\sim$}}\raise5.truept
\hbox{$>$}\ }}
\def\rho{\varrho}
\def\mdot{\dot M}

\begin{document}

\title{X--ray Emission from Old, Isolated, Accreting Neutron Stars}

\author{ R. TUROLLA }

\address{Dept. of Physics, University of Padova,  Via Marzolo 8, 35131 
Padova, Italy}


\maketitle\abstracts{Accretion onto old, isolated neutron stars
is reviewed. The detection of their X--ray emission with ROSAT is discussed
and a summary of the current status of observations is presented.}

\section{Introduction}

There is now a circumstantial evidence that most of the $\sim 600$ known
pulsars are born in the core collapse following a supernova explosion.
The estimated rate of SN events leading
to the formation of a NS is $\approx 10^{-2}$ yr$^{-1}$ at present,
which means that the Galaxy should contain $\approx 10^8-10^9$ neutron stars 
(NSs). Newly born NSs have large magnetic fields ($B\sim 10^{12}$ G), short 
periods ($P\mincir 0.1$ s) and show up as radio pulsars. A rough 
estimate of the life span of the pulsar phase, $P/\dot P\approx
10^7$ yr, shows that the present number of active pulsars is $\approx
10^5$ which is only a very tiny fraction of the estimated number of Galactic
NSs. Since the NS temperature is expected to drop from the initial 
$\sim 10^{11}$ K to $\mincir10^5$ K in $\sim 10^5$ yr, 
old, isolated NSs which have passed beyond the pulsar phase
are expected to be cold objects which do not emit any appreciable
amount of radiation.
However,
as it was suggested long ago in a pioneering paper by Ostriker 
{\it et al\/}~\cite{ors}, old, isolated NSs (ONSs) may accrete the interstellar 
medium and show up as weak, extreme UV/soft X--ray sources. The expected
spatial density of ONSs is 
$\approx 10^{-4} N_8 \ {\rm pc}^{-3}$, which implies that, on average, ONSs
are $\sim 15$ pc apart. Given their large number and relative proximity, the 
observation of isolated
accreting NSs may be already within reach. 

\section{Accretion Onto Isolated ONSs}

A spherical star (radius $R$, mass $M$) moving with velocity $v$ relative to 
an ambient medium of density $n$ accretes at the Bondi rate

\begin{equation}
\mdot \sim 10^{11}n v_{10}^{-3}M^2 \ {\rm g/s}\, .
\end{equation}
In the case
of NSs, however, the presence of a strong magnetic field and of fast rotation
may inhibit accretion because of the momentum outflow, produced by the 
spinning dipole, and of the propeller effect, induced by the corotating
magnetosphere~\cite{bm}~\cite{tcl}.

Accretion will occur only if the gravitational energy density of the incoming 
material 
exceeds the energy density of the relativistic outflow at the accretion 
radius. This condition is met only when the NS has spun down to a period
$P\magcir P_{crit}\sim 10 B_{12}^{1/2}\mdot_{11}^{-1/4}(r_A)_{14} R_{6}^{3/2}
M^{-1/8} \ {\rm s}\, .$
After $P$ has increased above $P_{crit}$ the infalling material proceeds
undisturbed until the NS magnetic energy density balances the matter 
bulk kinetic energy density at the Alfven radius. The 
corotating magnetosphere will then prevent the accreting material to go any 
further, unless the gravitational acceleration at the Alfven radius is larger
than the centrifugal acceleration which implies that 
$P\magcir P_A\sim 10^3B_{12}^{6/7}\mdot_{11}^{-1/2}M^{-1/2} \ {\rm s}\, .$

The timescale for spin--down by dipole radiation needed to meet the first
condition is a few Gyr,
so a large fraction of ONSs can have spun down sufficiently to overcome
the first barrier, but $P_A$ is so large that it can not be reached by
magnetic
dipole radiation. There are, however, two effects that may play an important 
role in slowing down NSs rotation, making accretion possible: the decay of the
B--field and the torque exerted by the accreting material on the NS itself. 
If the
magnetic field decays on a timescale $\approx 10^7$ yr, leaving possibly
a relic component of $\approx 10^8-10^9$ G, both 
$P_{crit}$ and $P_A$ are comfortably below the age of the Galaxy. Even for
a constant B--field of $\sim 10^{12}$ G, the torque could spin down the
NS in a time $\mincir 1$ Gyr~\cite{tcl}.

The accretion luminosity is

\begin{equation}
L\sim 2\times 10^{31}\mdot_{11} \ {\rm erg/s}\ll L_{Edd}
\end{equation}
which implies that
no energy is released before the flow hits the outermost stellar layers. Here
accreting protons are decelerated by Coulomb collisions with atmospheric
electrons and/or by plasma interactions and the flow stops after penetrating
a few Thomson depths in the NS atmosphere. The bulk kinetic 
energy of the infalling protons is then transformed into thermal energy at 
and finally converted into electromagnetic radiation~\cite{zs}~\cite{bsw}.

\section{The Emitted Spectrum}

Let us assume that thermal bremsstrahlung is the dominant mechanism
for producing photons. This means that the gas temperature in the atmosphere
should not differ too much from the effective temperature

\begin{equation}
T_{eff}=\left({L\over{4\pi R^2\sigma}}\right)^{1/4}\sim 3.4\times 10^5
L_{31}^{1/4}R_6^{-1/2} \ {\rm K}\, .
\end{equation}

The typical values of density and temperature are such that the free--free
optical depth is much larger than unity, so thermal equilibrium is 
established in the dense inner layers. 
Compton scattering is not expected to modify the spectrum because of the
relatively low Thomson depth and electron temperature; moreover the cyclotron 
line contribution to the total luminosity never exceeds
a few percent~\cite{nel}. 
The spectrum emitted by accreting ONSs can be then assumed to be a blackbody 
at $T_{eff}$, at least in the first approximation.
Emission is peaked at an energy

\begin{equation}
E\sim 3kT_{eff}\sim 100L_{31}^{1/4}R_6^{1/2} \ {\rm eV}
\end{equation}
and falls in the extreme UV/soft X--ray range.

Although for $B\sim 10^9$ G the magnetic field is not going to produce any 
major effect on
emission/absorption, its presence has an important consequence on the 
emitted spectrum. If a relic field $B = 10^9$ G is present, the flow is 
funneled onto the polar caps and the effective temperature is higher because
of the reduced emitting area
$(T_{eff})_{mag}
\sim 5 B_9^{1/7}\mdot_{11}^{-1/14}R_6^{5/28}M^{-1/28}(T_{eff})_{unmag}
\, .$

Detailed radiative transfer calculations~\cite{ztzt}
have shown that the emerging spectrum is not Planckian: it is harder
than a blackbody at $T_{eff}$ and the hardening becomes more pronounced as
the luminosity decreases. Since the hardening factor is $\sim 2$ at 
$L\sim 10^{31} \ {\rm erg/s}$, spectra emitted by magnetized accreting ONSs
are expected to peak at 

\begin{equation}
E\sim 10(3kT_{eff})\sim 1 L_{31}^{1/4}R_6^{1/2} \ {\rm keV}\, .
\end{equation}

\section{Observability of Old, Isolated NSs}

Emission in the $\sim$ 0.1--1 keV band, makes ONSs possible targets for 
detectors on board ROSAT and EUVE. The
maximum distance $d_{max}$ at which a star of given luminosity $L$ produces
a count rate above threshold turns out to be always larger than
$\sim 300$ pc, which means that ONSs are indeed within reach of ROSAT.

The total number of detectable ONSs in a given field is obtained integrating
the NS distribution function (DF)

\begin{equation}
N = \int \, dv\int_\Omega\, d\Omega\int_0^{d_{max}}f(r,l,b,v)\, dr\, .
\end{equation}
Unfortunately the NS DF at present is subject to various
uncertainties mainly related to our poor understanding of the velocity
distribution of pulsars at birth. Under reasonable assumptions, 
Treves and Colpi~\cite{tc} and Blaes and Madau~\cite{bm}
concluded that $\sim 6000N_9$ ONSs, located within 500 pc, should appear
in the ROSAT All Sky Survey. 

Given that accreting ONSs should indeed be detectable with ROSAT, and
possibly EUVE, the obvious question is: how to pick them up in the sea
of still unidentified, weak sources present in the surveys? Possible
criteria for sorting out good ONS candidates are: the lack of any optical
counterpart down to $m_v\sim 24$, soft X--ray spectra, extreme X to optical
flux ratio and correlation with the denser phases of the ISM. 
Relatively high
count rates are expected from ONSs accreting in the denser regions closer
to the Sun. In this 
respect, dense molecular clouds, $n\sim 100 \ {\rm cm}^{-3}$, 
seem to provide a very favourable environment
for observing ONSs~\cite{bm}~\cite{cct}~\cite{ztzct}.
Although the local interstellar medium is 
underdense and relatively hot, it contains
at least one region where $n \sim 1 \ {\rm cm}^{-3}$. Due to their vicinity
about 10 ONSs are expected to be detectable at the relatively flux
limit of 0.1 counts/s in the PSPC survey, $\sim 5\%$ of NOIDs at this 
limit~\cite{zztt}. 

Besides the detectability of individual sources, ONSs could also reveal
themselves through their contribution to the diffuse X--ray emission.
Zane {\it et al\/}~\cite{ztzct} have addressed this issue and concluded that 
magnetized, accreting ONSs can contribute up to $\sim 20 \%$ of the 
unresolved soft X--ray excess observed at high latitudes in the X--ray
background. Their number of resolved sources, $\sim 10 \ 
{\rm deg}^{-2}$,
although still consistent with the average density on NOIDs in PSPC
pointings ($\sim 30 \ {\rm deg}^{-2}$) would however imply that one
NOID in three should be an ONS. 
The Galactic Center, where
both the star and gas density is very high, could provide also an excellent
site for revealing ONSs emission. Assuming that ONSs are $\sim 1\%$ of
stars in the GC, Zane {\it et al\/}~\cite{ztt} were able to reproduce 
satisfactorily GRANAT data for the diffuse X--ray emission in the 2.5--7
keV band~\cite{smp}.

Theoretical predictions about the detectability of ONSs 
are rather optimistic, but do we actually observe them? Up to now there
have been only two claims of a possible detection of an accreting ONSs:
MS 0317.7-6647~\cite{sto} and RXJ 186535-3754~\cite{wwn}.
In both cases the source was detected with ROSAT PSPC and the evidences in 
favour of the ONS option (soft, blackbody
spectrum, lack of optical counterparts, positional coincidence with a cloud)
are strong but not completely compelling.
Very recently Belloni, Zampieri \& Campana~\cite{bzc} have performed a 
systematic search for ONSs in two molecular clouds in Cygnus, Rift and OB7, 
analyzing archive ROSAT PSPC pointings containing 109 sources. For 105 of 
them an optical counterpart
was identified, leaving 4 NOIDs with no counterpart
above $m_r\sim 20$. It should be stressed that the observed number of NOIDs
is $\mincir$ the predicted number of ONSs. 
A larger sample region in Cygnus, comprising a 
large part of the cloud OB7, has been investigated in detail by Motch
{\it et al\/}~\cite{mo}. using R--ASS data. The survey is complete at about
0.02 counts/s and, at this flux level, 
only 8 NOIDs are left. They do not seem to be correlated with the denser 
phases of
the ISM. Again this figure is lower than the estimated number of ONSs emitting
above 0.02 counts/s which is $\magcir 10$~\cite{ztzct}.

\section{Conclusions}

Although observations are still far from being conclusive, they seem to support
a number of detectable ONSs lower than what predicted by theory. There
are several effects that can be responsible for such discrepancy:
i) the total number of Galactic ONSs is $<10^8$, although this is
unplausible unless our understanding of SN rates is grossly wrong;
ii) accretion is inhibited because there is no magnetic field decay
or because of preheating in the accretion flow~\cite{bwm};
iii) ONSs are faster (and $\mdot$ is reduced) either as consequence of 
a higher velocity at birth~\cite{ll} or because of their dynamical 
interaction with the spiral arms~\cite{mb}.

\end{document}